\documentclass[%
reprint,
superscriptaddress,
showpacs,preprintnumbers,
nobibnotes,
 amsmath,amssymb,
 aps,
prb,
]{revtex4-1}
\usepackage[english]{babel}
\usepackage{graphicx}
\usepackage{dcolumn}
\usepackage{bm}
\usepackage{filecontents}
\usepackage{tabularx}
\newcolumntype{C}[1]{>{\centering\arraybackslash}p{#1}}

\usepackage{amsmath}

\usepackage{makecell}
\usepackage{cellspace}
\usepackage{hyperref}

\begin{document}
\hyphenation{Rehberg}


\title{Measuring magnetic moments of polydisperse ferrofluids \\ utilizing the inverse Langevin function}

\author{Ingo Rehberg}
\affiliation{Experimental Physics 5, University of Bayreuth, 95440 Bayreuth, Germany}

\author{Reinhard Richter} 
\affiliation{Experimental Physics 5, University of Bayreuth, 95440 Bayreuth, Germany}

\author{Stefan Hartung} 
\affiliation{Experimental Physics 5, University of Bayreuth, 95440 Bayreuth, Germany}

\author{Niklas Lucht} 
\affiliation{Institute of Physical Chemistry, University of Hamburg, 20146 Hamburg, Germany}

\author{Birgit Hankiewicz} 
\affiliation{Institute of Physical Chemistry, University of Hamburg, 20146 Hamburg, Germany}

\author{Thomas Friedrich} 
\affiliation{Institute of Medical Engineering, University of L\"{u}beck, 23562 L\"{u}beck, Germany}

\date{ \today}
\newcommand{\figwidth}{1\linewidth}
\begin{abstract}
The dipole strength of magnetic particles in a suspension is obtained by a graphical rectification of the magnetization curves based on the inverse Langevin function. The method yields the arithmetic and the harmonic mean of the particle distribution. It has an advantage compared to the fitting of magnetization curves to some appropriate mathematical model: It does not rely on assuming a particular distribution function of the particles.  
\end{abstract}
\maketitle

Ferrofluids, i.e.\,colloidal suspensions of magnetic particles, can be characterized by their magnetization curve, which reveals superparamagnetic behavior \cite{Rosensweig2013}. In particular, it is possible to obtain an estimate of the dipole moment distribution of the colloidal particles within the fluid from that curve 
\cite{Elmore1938}, which provides a convenient kind of magnetogranulometry \cite{Berkovski1996}. The underlying analysis of the magnetization curves is well defined for the case of small particle concentrations, where the interaction of the individual magnetic particles can be neglected.  
\begin{figure}[h!b]
\centering
\includegraphics[width=\figwidth]{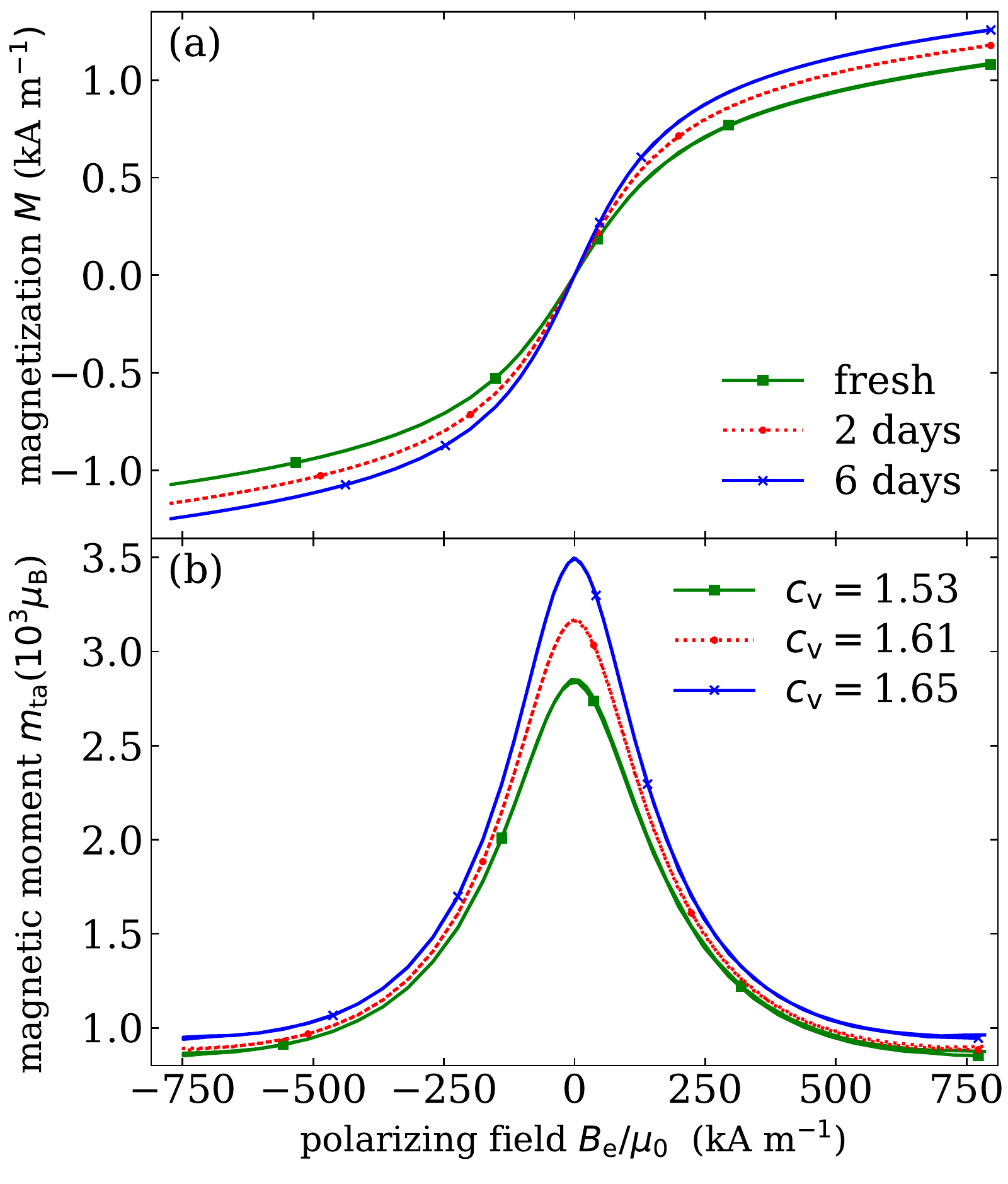}
\caption{Aging of a nanocube fluid. (a) The magnetization of a freshly prepared ferrofluid is presented together with one obtained two (six) days later. During the measurements, the magnetizing field strength went from about 750\,kA/m to -750\,kA/m and back within a period of 108\,minutes. The measurements are presented as polygonal lines, every 30th data point is shown to label them. (b) The curves shown in the lower part are derived from the magnetization curves and give information about the magnetic moments of the suspended particles. The maximum corresponds to the arithmetic mean $m_\mathrm{a}$, and the asymptotic value for large polarizing fields to the harmonic mean $m_\mathrm{h}$. The corresponding estimator for the coefficient of variation $c_\mathrm{v}$ is listed in the lower legend.}
\label{temp_evolution}
\end{figure}
The examination of the magnetization curves is thus a suitable tool to get an idea about the particle size distribution within the fluid, and in particular, it is suitable to resolve changes of the distribution, i.e.\ to monitor and characterize the aging of a colloidal suspension of magnetic particles. The extraction of the moment distribution function is done by assuming some continuous distribution function like, e.g., the gamma- or log-normal distribution with adjustable parameters. The distribution function is then obtained by fitting the corresponding magnetization curve to the measured one. Some examples, together with a critical comparison, are presented in Ref.\,\cite{Ivanov2007}. Alternatively, a distribution with discrete $\delta-$peaks can be assumed \cite{Mehdi2015,Rosenfeldt2018}. If no knowledge about the particle distribution function is available, an unprejudiced ansatz can be made in connection with a regularization scheme. This procedure yields at least reproducible results for the particle distribution function, an example is given in Ref.\,\cite{WeserStierstadt1985}. If the resulting distribution functions contain negative concentrations, additional mathematical insights are needed in order to interpret the results.

The computed magnetization curve is in the dilute limit a folding of the Langevin function --- which describes the magnetization of a sufficiently dilute monodisperse solution --- with the assumed particle size distribution function. 
For this kind of extraction procedure, the Langevin function has an unpleasant feature: The folding of different distribution curves with that function can give very similar, almost identical, results \cite{Martinet1983}. The situation is comparable to the method of extracting the characteristics of a polydisperse particle size distribution from the analysis of dynamic light scattering experiments, a prominent example for a mathematically ill-conditioned problem \cite{Berne2000}. The corresponding aspect of the Langevin function has been discussed in some detail by Potton et al.\cite{Potton1984}, who used a maximum entropy method to face the ensuing complications.

In this paper, we demonstrate a method which circumvents these difficulties by not even trying to obtain the complete distribution function. It is basically a graphical rectification of the magnetization curve and reveals important parameters of the magnetic moment distribution, but does not rely on assuming a particular distribution function of the magnetic particles. Our analysis of the rectified curves is, however, based on the limit of small concentrations. For larger concentrations, the interaction between the magnetic particles lead to additional complications \cite{Ivanov2007, Embs2007} which are not addressed to in the present paper.     
 
To give a motivation for the method, Fig.\,\ref{temp_evolution} provides an example of this rectification method to characterize an aging process of a ferrofluid. It makes use of data taken from the literature \cite{Mehdi2015,Rosenfeldt2018} describing the formation of magnetic clusters in a colloidal suspension of nanocubes. They characterize the aging of cubic nanoparticles (8\,wt\%, iron oxide, edge length 9\,nm) in solution triggered by a magnetic field (800\,kA/m for 4\,h). Figure \ref{temp_evolution}(a) shows magnetization curves of that fluid for three different times. They were obtained with a vibrating sample magnetometer described in detail by Friedrich et al.\cite{RevSciInstr-83-045106-2012}. The first data set was obtained for a relatively fresh sample, which had been exposed to a magnetizing field of about 800\,kA/m for four hours. The magnetization curves in Fig.\,\ref{temp_evolution}(a) show an increasing slope with the time elapsed. This aging process is interpreted as the manifestation of the clustering of the magnetic particles.
\begin{figure}[h!]
\centering
\includegraphics[width=\figwidth]{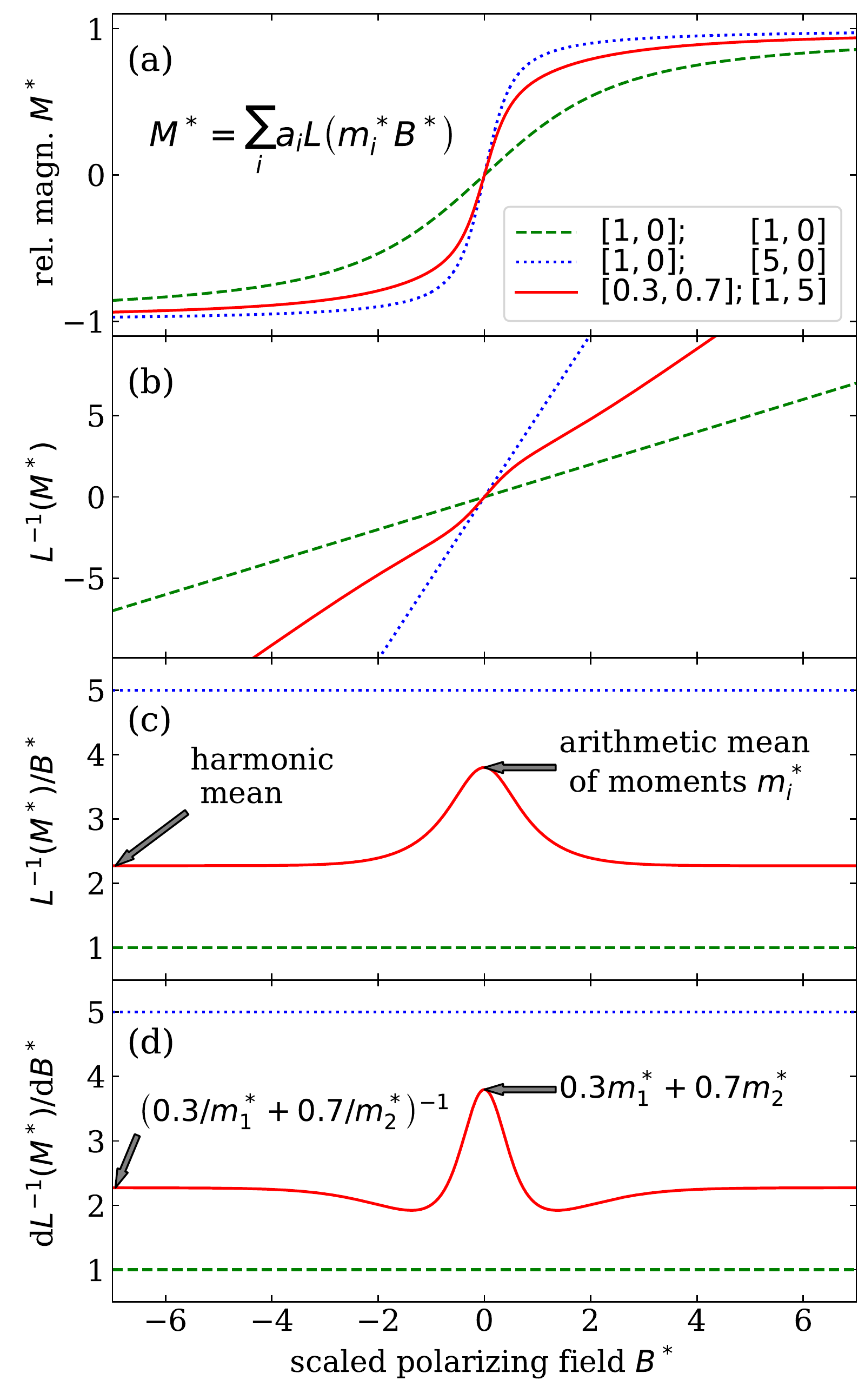}
\caption{The data processing demonstrated by three artificial magnetization curves. (a) The magnetization curves of two monodisperse (dashed and dotted lines) and a bidisperse solution. The first pair of numbers in the legend re\-presents the relative fraction $a_{1}$ and $a_{2}$ , and the second one the corresponding magnetic moments $m_{1}$ and $m_{2}$. (b) The inverse Langevin function $\mathrm{L}^{-1}$ of the relative magnetization. The straight dashed and dotted lines correspond to the two monodisperse distributions, the slightly curved solid line to the bidisperse distribution. (c) The chord slope of the rectified curves. The monodisperse distributions lead to constant values (dashed and dotted lines) which represent the strengths of the magnetic dipole moment. The bidisperse curve yields the arithmetic mean of the two contributing moments as its maximum value, and the harmonic mean as the asymptotic value for large polarizing fields. (d) The tangential slope of the $\mathrm{L}^{-1}(M^*)$ curves.}
\label{explanation}
\end{figure}
Some features of the change of these curves can be seen more clearly in Fig.\,\ref{temp_evolution}(b). Here, the appropriately scaled slope of the inverse Langevin function $\mathrm{L}^{-1}$ of the magnetization data has been plotted. The ensuing curves yield the arithmetic mean of the dipole distribution at its center, and the harmonic mean as the asymptotic value for large polarizing fields. 

To explain this, we illustrate the data processing by artificial magnetization curves in Fig.\,\ref{explanation}. A monodisperse dilute solution of particles with a magnetic moment $m$ is expected to be described by a magnetization 
\begin{equation*}
M=M_\mathrm{s} \mathrm{L}\left( \frac{m B}{k_\mathrm{B}T} \right),\  \mathrm{with}\ \mathrm{L}(x)=\mathrm{coth}(x)-\frac{1}{x} 
\end{equation*}
In Fig.\,\ref{explanation}(a), the abbreviations 
\begin{equation*}
M^*=M/M_s,\  m^*=m/\mu_\mathrm{B},\  \mathrm{and}\ B^*=B\frac{\mu_\mathrm{B}}{k_\mathrm{B}T} 
\end{equation*}
are used. It displays the magnetization of two monodisperse fluids with $m^*=1$ and $m^*=5$, respectively,  and one for a bidisperse 30\%/70\%- mixture. All three curves show a fairly similar shape. To bring out the difference between these curves more clearly, it helps to take the inverse Langevin function  $\mathrm{L}^{-1}(M^*)$ as shown in Fig.\,\ref{explanation}(b). The two monodisperse curves reveal a constant slope --- in this sense the magnetization curve is rectified --- while that of the mixture appears slightly more complicated. To bring out these differences quantitatively, both the chord slope $m^*_\mathrm{ch}=\frac{\mathrm{L}^{-1}}{B*}$ or the tangential slope $m^*_\mathrm{ta}=\frac{\mathrm{dL}^{-1}}{\mathrm{d}B*}$ can be used to obtain a value for what can be called an "effective magnetic moment". $m^*_\mathrm{ch}$ is shown in Fig.\,\ref{explanation}(c) and the tangential slope $m^*_\mathrm{ta}$ in Fig.\,\ref{explanation}(d). In both cases, the monodisperse curve yields the constant value $m^*$, which is proportional to the magnetic moment of the particles. 
 
The more interesting part is the interpretation of the non-constant curves obtained for the bidisperse mixture. Both methods yield the same maximum in the center, i.e.\,for the magnetizing field $B^*=0$. Near this point $\mathrm{L}(B^*m^*)\approx \frac{B^*m^*}{3}$, thus the derivative represents the appropriately weighted sum of the two slopes of the monodisperse magnetization curves, i.e.\ the arithmetic mean $m^*_\mathrm{a}=\left<m^*_i\right>$ of the magnetic moments involved. Its value is $0.3 m_1+0.7 m_2= 3.8$ for this particular example.

Both methods also yield the same results for large values of $B^*$. For the interpretation of this value, one has to recall that the Langevin function converges to its asymptotic value, 1, like $1/(B^* m^*)$, which means that the slope is inversely proportional to the magnetic moment. Consequently, the slope for the bidisperse curve can be obtained by the weighted sum of the inverse moments, the harmonic mean $m^*_\mathrm{h}=\left<1/m^*_i\right>^{-1}$. It is $\left(0.3/m_1+0.7/m_2\right)^{-1}\approx 2.27 $ for this example. 

Whether the chord slope or the tangential slope should be used to obtain the 
effective magnetic moment for real data is a practical issue. When dealing with a poor signal/noise ratio, data obtained from the chord slope have the advantage to show less scatter. On the other hand, the effective magnetic moments obtained from the tangential slope have the advantage to converge faster towards the asymptotic limit, which is important when the scaled applied field $B^*$ is still far from the saturation field. A practical value for judging the strength of the polarizing field could be given by that field where the magnetization reaches 90\% of $M_\mathrm{s}$. The value for the corresponding polarizing field is then given by $L(m^*B^*)=0.9$,  yielding $B^*=L^{-1}(0.9)/m^*\approx 10.0/m^*$.

The difference between the arithmetic and the harmonic mean values, $m_\mathrm{a} - m_\mathrm{h}$,  can be taken as a direct order parameter for the amount of polydispersity: It is zero for a monodisperse distribution and increases with the width of the distribution. In fact, this difference divided by the harmonic mean provides an estimator for the relative standard deviation (RSD, also called coefficient of variation $c_\mathrm{v}$). More precisely, we obtain the coefficient of variation as $c_\mathrm{v}=\sqrt{\frac{m_\mathrm{a}-m_\mathrm{h}}{m_\mathrm{h}}}$. Additionally, the square root of their product yields an estimator for the geometric mean $m_\mathrm{g}=\sqrt{m_\mathrm{a}  m_\mathrm{h}}$. However, these last two statements are only correct for certain distribution functions of the magnetic moment, including the log-normal distribution, which seems to be the most prominent one assumed within the granulometric analysis of magnetization curves. 

\begin{figure}[hb]
\centering
\includegraphics[width=\figwidth]{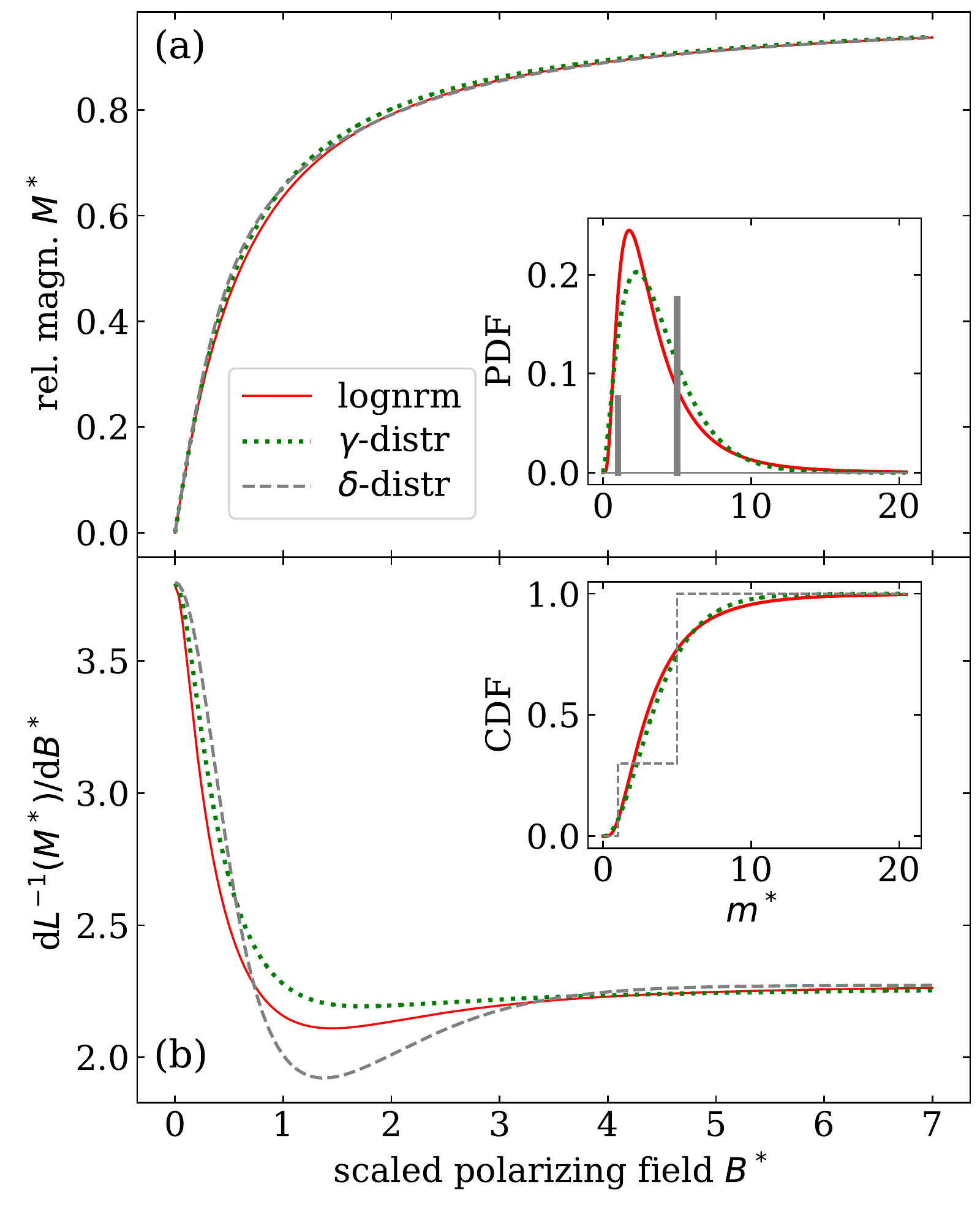}
\caption{A comparison between magnetization curves calculated for the bidisperse distribution with two $\delta$-peaks introduced in the example in Fig.\,\ref{explanation} (dashed gray line), the log-normal distribution (solid red line), and the $\gamma$-distribution (dotted green line). The parameters are chosen such that all three distributions have the same values of the harmonic and the arithmetic mean. Therefore, all curves in (b) start for $B^*=0$ at the same value of 3.8 and approach the value of 2.27 for high field strength. The inset (a) shows the partial and inset (b) the cumulative distribution functions of the three distributions.}
\label{figure3}
\end{figure}

To illustrate the procedure with more realistic distributions than the artificial bidisperse one used in Fig.\,\ref{explanation}, we compare this bidisperse distribution with a suitably chosen log-normal and gamma distribution \cite{Ivanov2007}. More precisely, in both cases we chose that distribution which has the same arithmetic and harmonic mean as the bidisperse one. This is possible because both functions contain two adjustable parameters. The comparison is presented in Fig.\,\ref{figure3}. The inset of Fig.\,\ref{figure3}(a) shows the distribution function for the three cases. The continuous functions are the log-normal and gamma distribution, while the bidisperse distribution function is basically zero, except for the two $\delta$-peaks. The corresponding cumulative distribution functions for the three examples are shown in the inset of Fig.\,\ref{figure3}(b).

Note that in spite of the drastically different distribution functions, the corresponding magnetization curves displayed in Fig.\,\ref{figure3}(a) are almost non-distinguishable. This is an exemplary illustration of the ill-conditioned nature of magnetogranulometry mentioned in the introduction. 

Taking the derivative of the inverse $\mathrm{dL}^{-1}(M^*)/\mathrm{d}B^*$ helps to bring out the differences in the three magnetization curves more clearly, as shown in Fig.\,\ref{figure3}(b). More importantly, this effective magnetic moment $m^*_\mathrm{ ta}$ reveals the correct arithmetic and harmonic mean for all three distribution functions, as expected. 

\begin{figure}[ht]
\centering
\includegraphics[width=\figwidth]{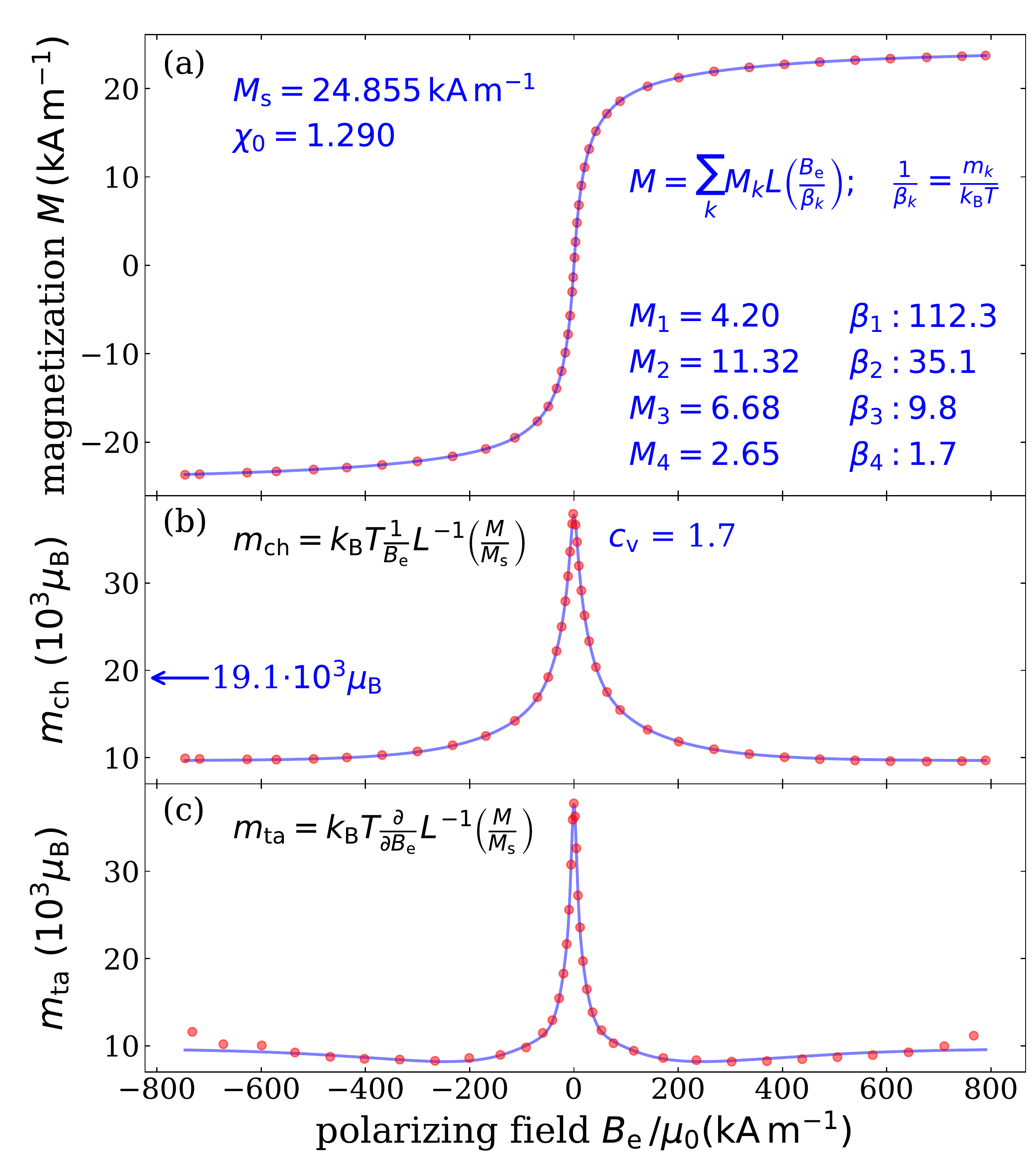}
\caption{The method illustrated by the the commercially available ferrofluid EMG909. (a) The measured magnetization curve (red dots, only every 10th data point is shown) is fitted by a superposition of four Langevin functions (solid blue line) indicated by the $M_k$ given in $\mathrm{kAm^{-1}}$. The corresponding $\beta_k$ yields the magnetic moment $m_k$ provided in $\mathrm{k\mu_B}$. The resulting saturation magnetization $M_\mathrm{s}$ and the initial susceptibility $\chi_\mathrm{0}$ are listed as well. (b) The effective magnetic moment $m_\mathrm{ch}$ obtained from the data (red dots) and the fitting function (solid blue line). The $c_\mathrm{v}$ obtained from the arithmetic and harmonic mean of the magnetic moments is listed, and the blue arrow points to the value of the corresponding geometric mean. (c) The effective magnetic moment $m_\mathrm{ta}$ obtained from the data (red dots) and the fitting function (solid blue line)}
\label{EMG909}
\end{figure}

Finally we would like to illustrate the method by analyzing magnetization curves of two additional samples of ferrofluids. The one measured for commercially available EMG909 (EMG909, Lot H030308A, Ferrotec) is presented in Fig.\,\ref{EMG909}(a).
The "polarizing field" used for the horizontal axis is the field acting on a magnetic particle. We used the lowest order to determine that field, namely the Weiss correction $H_\mathrm{e}=H_\mathrm{i}+M/3$, see e.g.\,Ref.\cite{Ivanov2007} for a discussion of this correction. Note that in our case the correction term $M/3$ exactly cancels out the demagnetization factor provided by our spherical sample holder, leading to $H_\mathrm{e}=H_\mathrm{0}$, and $B_\mathrm{e}=B_\mathrm{0}$. Thus, in our case the polarizing field $B_\mathrm{e}$ turns out to be the one measured far from our magnetized sphere, $B_\mathrm{0}$. Note that the resulting plot --- with the effective $B_\mathrm{e}$-field used for the x-axis --- is slightly different from the more common practice, where the inner magnetic field $H_\mathrm{i}$ is used for the horizontal axis of the magnetization curve. For the latter kind of plot, however, taking $\mathrm{L}^{-1}(M/M_\mathrm{s})$ would not produce a straight line even for a monodisperse ferrofluid. This would make the rectification method proposed here less powerful.

The measured magnetization data can well be represented by a superposition of four Langevin functions
\begin{equation*}
M(B_\mathrm{e})= \sum_{k=1}^{4} M_k \mathrm{L}\left(\frac{B_\mathrm{e}}{\beta_k}\right), \mathrm{with} \frac{1}{\beta_k}=\frac{m_k}{k_\mathrm{B}T}. 
\end {equation*}
This $M(B_\mathrm{e})$ resulting from this "quad-disperse" distribution function provides a convenient fitting curve for the magnetization data, with the $M_k$ and $\beta_k$ as fit parameters, and is shown as a solid line in the upper part. It serves primarily for giving a smooth and analytic representation of the data. In addition, it can be used to calculate the so called Langevin susceptibility $\chi_\mathrm{L}$ as the slope of the magnetization curve in its origin. From $\chi_\mathrm{L}$, the initial susceptibility $\chi_0=\frac{\mathrm{d}M}{\mathrm{d}H_\mathrm{i}}$ is obtained as $\chi_0=\frac{\chi_\mathrm{L}}{1-\chi_\mathrm{L}/3}$, which is provided in the figure as well. While this number is an important characteristic number for ferrofluids in general, its value is not needed for the further analysis presented here, but it helps to label the fluid and to judge its concentration. The saturation magnetization can be obtained from the fitting parameters as $M_\mathrm{s}=\sum_{k=1}^{4} M_k$. 

Fig.\,\ref{EMG909}(b) shows the effective magnetic moment $m_\mathrm{ch}$ obtained from the chord slope. The red dots are obtained directly from the data. The solid blue line stems from the fit to the magnetization curve. Both numbers agree fairly well. Note that there is a small asymmetry with respect to the y-axis within the data, which the ansatz for the quad-disperse fitting function cannot produce.

These small differences between the data and the fitted curve can be seen more clearly in Fig.\,\ref{EMG909}(c), where the effective magnetic moment $m_\mathrm{ta}$ is shown. But even here the signal/noise ratio seems good enough to extract the numbers for $m_\mathrm{a}$ and $m_\mathrm{h}$, and the corresponding guesses for the geometric mean $m_\mathrm{g}$ and the relative standard deviation $c_\mathrm{v}$.  

\begin{figure}[hb]
\centering
\includegraphics[width=\figwidth]{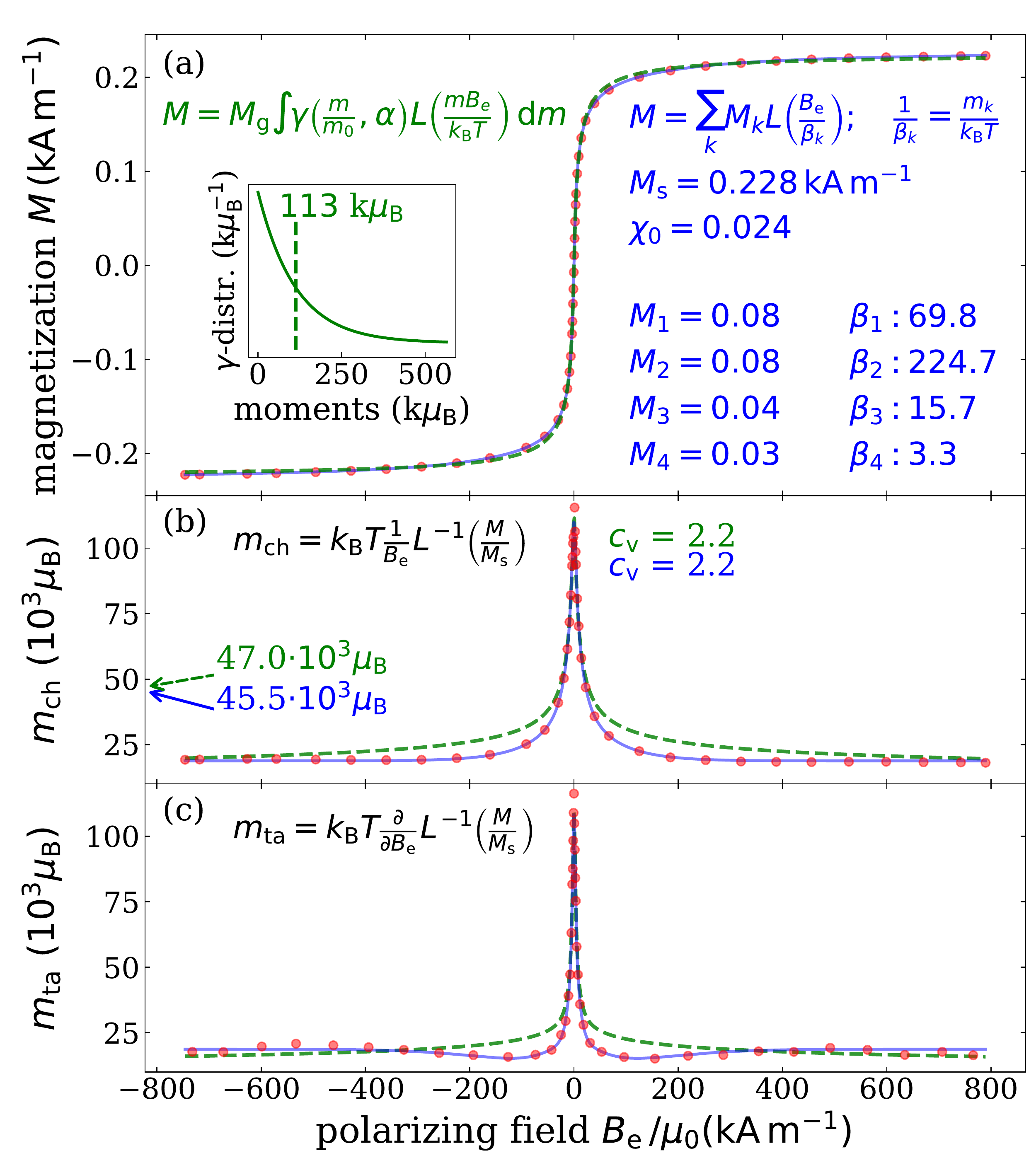}
\caption{The method illustrated by a $\mathrm{CoFe}_2\mathrm{O}_4$-ferrofluid. The features are the same as explained in Fig.\,\ref{EMG909}, and in addition, a fit to a $\gamma$-distribution (solid green line) shown in the inset has been performed here. While the differences of the two fitting functions in (a) are barely visible, (b) and (c) bring out these tiny differences more clearly. The geometric mean of both fits is indicated by the arrows in (b).}
\label{CoFe}
\end{figure}

For demonstrating the method also with a different chemical species, we use a cobalt-ferrite-based ferrofluid. It was synthesized in a one-step process with a subsequent stabilization step after a modified synthesis procedure of Nappini et al.\,\cite{Nappini2015}. For the synthesis both iron and cobalt salts were precipitated in a boiling solution of sodium hydroxide. The particles were magnetically separated by holding a permanent magnet (with a surface field of about 1\,T and a diameter of about 3\,cm) onto the reaction vessel for a few minutes and rinsed with water. This step was repeated until a neutral pH value was reached, typically about three times, then the particles were stabilized in a sodium citrate solution. The resulting magnetization curve is shown in Fig.\,\ref{CoFe}(a). It can also fairly precisely be fitted by assuming a quad-disperse solution, as shown by the blue line. In addition, we have also fitted a $\gamma$-distribution here, as advocated in \cite{Ivanov2007}. The resulting distribution is shown in the inset. The corresponding magnetization shown by the green line fits the data almost as good as the quad-disperse one, which is just considered as another manifestation of the ill-posed character of this inverse problem. 

Displaying the resulting magnetic moments in Fig.\,\ref{CoFe}(b) and (c) brings out the differences between the two magnetization curves more clearly. It also reveals that the quad-disperse fit is closer to the data, which is no surprise, because that fit contains eight fitting parameters, while the $\gamma$-distribution only has two. With a relative standard deviation of $c_\mathrm{v}=2.2$, the distribution function of the $\mathrm{CoFe}_2\mathrm{O}_4$-ferrofluid is wider compared to the EMG909 fluid presented in Fig.\,\ref{EMG909}. That might be a manifestation of the fact that our fluid was relatively freshly prepared, and no special measures were taken in order to obtain a more monodisperse solution. On the other hand, special measures to obtain monodispersity were taken for the fluid analyzed in Fig.\,\ref{temp_evolution}, which contained originally fairly monodisperse nanocubes. Here the monotonic increase of $c_\mathrm{v}$ with time is interpreted as a result of the formation of supercubes \cite{Mehdi2015,Rosenfeldt2018}.

In summary, we have demonstrated the use of a graphical rectification method revealing the characteristic magnetic moments of the particles in a ferrofluid from their magnetization curves. In particular, the arithmetic and the harmonic mean of the moments, $m_\mathrm{a}$ and $m_\mathrm{h}$, can be read off from a plot of the effective magnetic moment. The method works without the need to assume a specific distribution function, thus circumventing the difficulties stemming from an ill-posed problem for the interpretation of those functions. As secondary results, the method yields a guess for the relative standard deviation $c_\mathrm{v}$ and the geometric mean $m_\mathrm{g}$, although that guess can strictly be justified only for certain distributions including the log-normal one. The method applied here can be justified for dilute solutions, higher order corrections for larger concentrations \cite{Ivanov2007, Embs2007} have not been taken into account. A corresponding graphical method for the examination of light scattering data in terms of granulometric information is currently under investigation. 

The open source Python code for the graphical display of the magnetization curves together with the ensuing magnetic moments is still under construction, but we are happy to provide the current version on request.

It is a pleasure to thank H.\,R.\,Brand for stimulating discussions and suggestions.

\bibliography{Rehberg19}

\end{document}